\documentclass{ws-p9-75x6-50}

\begin{document}

\title{Binary black holes coalescence: transition from adiabatic 
inspiral to plunge \footnote{Contributed paper to the {\em  {\rm IX} Marcel Grossmann Meeting} in Rome, July 2000.}}

\author{Alessandra Buonanno}

\address{Theoretical Astrophysics and Relativity Group \\
California Institute of Technology, Pasadena, 
CA, 91125, USA\\E-mail: buonanno@tapir.caltech.edu}

\author{Thibault Damour}

\address{Institut des Hautes Etudes Scientifiques, 91440
Bures-sur-Yvette, France 
\\E-mail: damour@ihes.fr}

\maketitle

\abstracts{Using two recent techniques giving 
non-perturbative re-summed estimates of the damping 
and of the conservative part of the dynamics of two-body systems, 
we describe the transition between adiabatic 
inspiral and plunge in binary non-spinning 
black holes moving along quasi-circular orbits.} 

It seems likely that the first detection of gravitational waves with 
the LIGO/VIRGO/... interferometers  will come from 
binary systems made of massive black holes of comparable masses,
say with a total mass
 $M \simeq 15 M_{\odot} + 15 M_{\odot}$. For such massive systems 
the gravitational-wave  frequency at the last 
stable (circular) orbit (LSO) is very close to the location 
of the minimum of the detector's noise spectral density. 
Therefore, for data analysis purposes it is quite desirable 
to have a thorough knowledge of the late dynamical evolution 
of comparable-mass binaries. 

When the two black holes become closer than $ \sim 10 M$, 
the post-Newtonian (PN) expansion, which adequately describes 
the motion at large separations, cannot be trusted any more, and 
non-perturbative techniques should be used to deal 
with the non linearities and the strong curvature effects 
of the final phase of evolution. 
Despite the progress made by the numerical relativity 
community during the last years,  
an estimate of the complete waveform emitted 
by a comparable-mass black hole binary has not been 
provided, yet.
Hence, to tackle this issue we have recently introduced 
a new \emph{ non-perturbative analytical} approach to  the
general relativistic two-body 
problem \cite{BD1}$^,$ \cite{BD2} which should be able to capture 
the crucial features of the transition from the 
adiabatic phase to the plunge.

Our approach combines two PN re-summation techniques. 
The first method~\cite{BD1} allows one to derive 
a non-perturbative estimate of the conservative 
part of the nonlinear force law determining the motion of 
comparable-mass binaries. The basic idea \cite{BD1} 
is to map the conservative {\em real} two-body dynamics 
(up to 2PN order) onto an {\em effective} one-body one, where a test 
particle of mass $\mu \equiv m_1 \, m_2 /M$ moves
in some effective background metric $g_{\mu \nu}^{\rm eff}$. 
As long as radiation reaction effects are 
not taken into account, the effective metric 
is just a deformation of the Schwarzschild 
metric with deformation parameter $\nu = \mu/M$. 
The effective description provides a way of re-summing in a non-perturbative 
manner the badly convergent PN-expanded dynamics 
of the real description, giving the following improved real Hamiltonian \cite{BD2}:
\begin{equation}
{H_{\rm real}^{\rm improved}} = 
M\,c^2\,\sqrt{1 + 2\nu\,\left ( \frac{{H_{\rm eff}^{{\nu}}} 
- \mu\,c^2}{\mu\,c^2}\right )}\,.
\label{eq1}
\end{equation}
The second non-perturbative technique \cite{DIS1} concerns 
radiation reaction effects, and gives  a re-summed 
estimate (using Pad\'e approximants) of the energy loss rate 
along circular orbits,
$\Phi_{\rm circ}$ (up to 2.5 PN order).
Combining the two re-summation 
methods yields a system of ordinary differential equations, 
which in spherical coordinates $(\varphi, R, P_{\varphi},P_R)$ reads \cite{BD2}:
\begin{equation}
\frac{d R}{d t} = 
\frac{\partial H^{\rm impr}_{\rm real}}{\partial P_R}\,, \quad 
\frac{d P_R}{d t} + \frac{\partial H^{\rm impr}_{\rm real}}{\partial R}= 0\,, 
\quad \frac{d \varphi}{d t} = 
\frac{\partial H^{\rm impr}_{\rm real}}{\partial P_\varphi}\,, \quad 
\frac{d P_\varphi}{d t} = -\frac{\Phi_{\rm circ}}{\dot{\varphi}}\,.
\label{eq2}
\end{equation}
When working linearly in the radial velocity $\dot{R}$,
the following characteristic equation describing the transition 
from the adiabatic inspiral to the plunge was derived \cite{BD2}:  
\begin{equation}
\frac{d^3 R}{d t^3} +  {\omega_R^2(R)}\,\frac{d R}{d t} = - {B_R(R)}.
\label{eq3}
\end{equation}
The quantity $\omega_R^2$ plays the role of a ``restoring force''. It is  
the square of the frequency of radial oscillations 
and  is proportional to the curvature 
of the effective radial potential (it vanishes
at the LSO). The quantity $- B_R$ ($\propto \nu$) 
is a ``driving force'', coming from gravitational 
radiation damping. The term $d^3 R/d t^3$ is an
``inertia term'', which is neglected in the adiabatic approximation, 
but should be retained when describing the motion in proximity 
of the LSO and beyond it. An equation analogous to Eq.~(\ref{eq3}) 
has also been independently derived recently \cite{OT} for a 
test particle moving along quasi-circular equatorial orbits 
in Kerr spacetime. Our approach has been recently extended to the
3PN level \cite{DJS}.

Using Eq.~(\ref{eq2}) and 
the canonical mapping between the effective and the real description 
\cite{BD1}, we can: (i) provide initial 
dynamical data, for numerical relativity investigations, 
for black holes that have just started their plunge motion 
and (ii) give, for data analysis purposes, an estimate 
of the gravitational waveform emitted throughout the inspiral, 
plunge and ring-down phases (see \cite{DIS3} for the data analysis
consequences of this waveform).

Let us finally mention two interesting facts: (1) the energy emitted 
during the plunge phase is only around $0.7 \% M$, with a comparable
energy loss $ \sim  0.7 \%  M$ 
during the following ring-down phase, and (2) the
phasing of the gravitational wave signal is essentially unchanged
if one turns off the radiation reaction term in Eq.~(\ref{eq2}),
 after LSO crossing.


\begin{thebibliography}{99}

\bibitem{BD1} A. Buonanno and T. Damour, Phys. Rev. D {\bf 59}, 084006 (1999). 
\bibitem{BD2} A. Buonanno and T. Damour, Phys. Rev. D {\bf 62}, 064015 (2000). 
\bibitem{DIS1} T. Damour, B.R. Iyer and 
B.S. Sathyaprakash, Phys. Rev. D {\bf 57}, 885 (1998).
\bibitem{OT} A. Ori and K.S. Thorne, [gr-qc/003032].
\bibitem{DJS} T. Damour, P. Jaranowski and G. Sch\"afer, 
Phys. Rev. D {\bf 62},  084011 (2000).
\bibitem{DIS3} T. Damour, B.R. Iyer and 
B.S. Sathyaprakash, [gr-qc/0010009].
\end{thebibliography}
\end{document}